# The potential fluctuation and its interfacial phenomena in molecular, micro, macro, and cosmic flow instabilities

Wei Li
Professor and Fellow of ASME
Department of Energy Engineering, Zhejiang University
Yuquan Campus, Hangzhou City, 310027, PR China.
Email: weili96@zju.edu.cn

**Abstract**

Flow instabilities play important roles in a wide range of engineering, geophysical, and astrophysical flows, ranging from supernova explosion in Crab Nebula, formation of clouds in sky, waves on ocean, to inertial confinement fusion capsules, making fusion energy a viable alternative energy source in the future. The potential for life is directly related to flow instability mixing as well. Previous researchers have focused on developed stages of flow instabilities by assuming SINE wave interface between fluids in the flow instabilities. No scientific research has been reported to investigate the origin of flow instabilities. The paper advances a new physics concept, potential fluctuation in flow based on the conservation of mass, which presents potential oscillatory SINE wave surface in the spatial and temporal dimensions at the interface of flow instabilities. Potential fluctuation is decided by the two densities and velocities in the flow as indicated by the relation of continuity. Even before the flow instabilities start to develop, potential fluctuation has already internally existed in flow. It is only decided by the densities and velocities of the two moving fluids and is not related to the surface topography of the boundary of flows in the flow instability. It is the 'gene' of flow instability. The paper presents breakthrough of understanding of micro, macro, and cosmic flow instabilities.

**Introduction**

Flow instabilities occur at the interfaces between moving fluids. Much effort has been expended over the past 150 years, beginning with the seminal work of L. Rayleigh, along with its modern development by G.I. Taylor, to predict the evolution of the instabilities and of the instability-induced mixing layers. It is known as the Rayleigh-Taylor instability (RTI). RTI happens occurs whenever a heavy fluid of density is supported against a gravity by a lighter fluid of density. Richtmyer-Meshkov instability (RMI) is the impulsive-acceleration limit of RTI. RMI arises when a shock passes through an interface between two fluids; RMI happens when the acceleration is toward either side of the interface, whereas for the RTI case one has instability only for acceleration in the direction of the lighter fluid and largely during a time where the external forcing is zero since the driving is impulsive. Kelvin-Helmholtz instability (KHI) occurs whenever velocity shear presents within a continuous fluid or when there is sufficient velocity difference across the interface between two fluids. KHI is the reason for the evolution of the mushroom structures during the RTI and RMI nonlinear process. RTI, RMI, and KHI are collectively referred to flow instabilities.

Flow instabilities play important roles in almost all scientific research areas related to flow. Some research areas are as following: two-phase flow, mixing in supersonic flows [1-4], heavy-nuclei collision [5], sonoluminescence [6-8], buoyancy-driven mixing [9], magnetically confined plasma [10, 11], quantum plasma [12-14], dusty plasma [15-18], Earth's ionosphere [19], oceanography [20], biology, solar atmosphere [21], climate dynamics [22, 23], weather inversions [24], lightning return stroke and natural discharge in the lower atmosphere [25-28], interaction between surface loads and internal planetary dynamics [29], filamentary structure on the Sun [30-32], galaxy clusters [33-40], Jupiter's and Saturn's magnetospheres [41], accretion onto the magnetospheres of neutron stars[42, 43], batholiths[44], convective thinning of the lithosphere [45], biogenesis, geophysical formations of volcanic islands [46-50], interaction between solar wind plasma and magnetospheric plasma at the Earth [51], salt tectonics [52], solid/liquid and solid/solid media [53-63], shock–magnetosphere interactions [64,65], and so on.

Flow instabilities have found many industrial applications as well. Some applications are as following: air conditioning system, steam turbine generator system, heat exchangers and internal combustors [66-68], inertial confinement fusion (ICF) capsules making fusion energy a viable alternative energy source, where high density shells are decelerated by low density fuel, premixed combustion [69-72], micron-scale fragment ejection [73-77]. industrial coating with thin liquid films [78], radiation pressure acceleration of heavy ions from laser-irradiated ultrathin foils [79], discharge from stacks and pollutant dispersion [80-82], reactive systems [83-85], expanding waves in supersonic flows [86], sediment transport [87], underwater explosions [88-91], relativistic jets and the surrounding medium [92], laser–material interactions with microfluid dynamics [93-95], acceleration of an elastic–plastic solid [96], chemical explosive into dilute aluminum particle clouds [97]; aerosol [98], energy loss in pipeline system [99], aluminum plates driven by detonation [100]; elastic–plastic media and solids [101-109], non-premixed reactive flows [110-113], compressed gas encased in a finite-sided cylindrical container [114] and so on. For an example, a review article in Nature Physics addresses that flow instabilities are important issues in the design of targets for laser fusion to achieve conditions under which inertial confinement is sufficient for efficient thermonuclear burn. Thermonuclear ignition is one milestone in the development of fusion energy, as well as a major scientific achievement with important applications in national security and basic sciences. The US and other countries have invested in large facilities in National Laboratories (for examples: Sandia National Laboratory, Los Alamos National Laboratory, Lawrence Livermore National Laboratory in US; Rutherford-Appleton Laboratory in UK, Russian Federal Nuclear Center, ShengGuang-II and III laser facility in China) to pursue it. Since 2012, progress has been made towards ignition conditions by using pulse shapes that are resistant to hydrodynamic instabilities over a range of ablator materials, pulse shapes and hohlraum gas fills, major challenges still remain to meet theoretical/simulation expectations [115].

Flow instabilities present triggering events that lead to fluid mixing in electro-hydro-dynamical and many other industrial processes. The slightest initial perturbation at the interface leads to tangential acceleration and will be amplified by material flowing down under influence of forces, which is magnified into a variety of interpenetrating structures. Previous studies focus on the influence of forces, for instance, by surface tension, elasticity, ablation, viscosity, accretion, plasticity or to any other effect causing forces on the interface. Because the topics presented here have exceptional practical and fundamental importance, thousands papers have been published to address the interpenetrating structures from every aspect. An example of inter-penetrating Rayleigh-Taylor fingers is supernova explosion in the Crab Nebula taken by NASA's Hubble Space Telescope. Different elements that were expelled during the explosion are shown in a variety of color. The filaments in the outer part of the nebula representing neutral oxygen, the singly-ionized sulfur, the doubly-ionized oxygen, and filaments which are the tattered remains of the star and consist mostly of hydrogen are shown in blue, green, red, and orange, respectively in Figure 1.

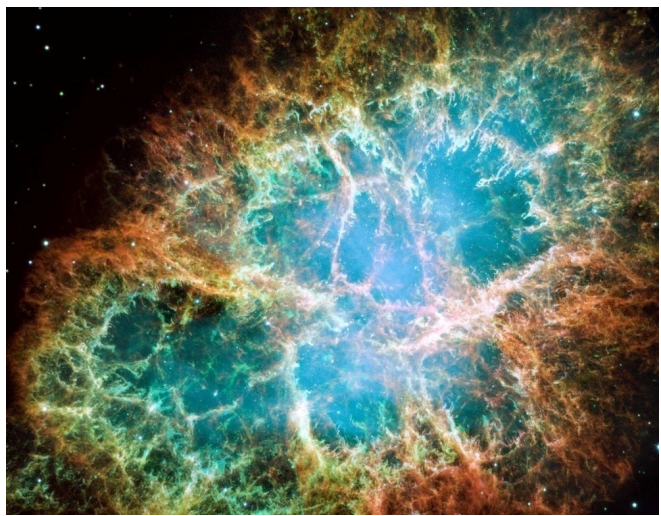

Figure 1    The expanding pulsar wind nebula powered by the Crab pulsar is sweeping up ejected material from the supernova. As the RTI develops, downward-moving irregularities are quickly magnified into sets of inter-penetrating Rayleigh-Taylor fingers [Wikipedia-http://en.wikipedia.org/wiki/Rayleigh-Taylor_instability] [116].

Inertial confinement fusion (ICF) determines the minimum energy required for ignition which is a major concern in making fusion energy a viable alternative energy source. RTI presents a serious design challenge for ICF capsules. The spherical shell filled with low-density gas is composed of an outer region, which forms the ablator, and an inner region of frozen or liquid deuterium tritium (DT), which forms the main fuel as shown in Figure 2. High density shell is decelerated by low density fuel in fusion capsules depending on the ratio of shell radius to thickness. Energy from a driver is delivered rapidly to the ablator, which heats up and expands.

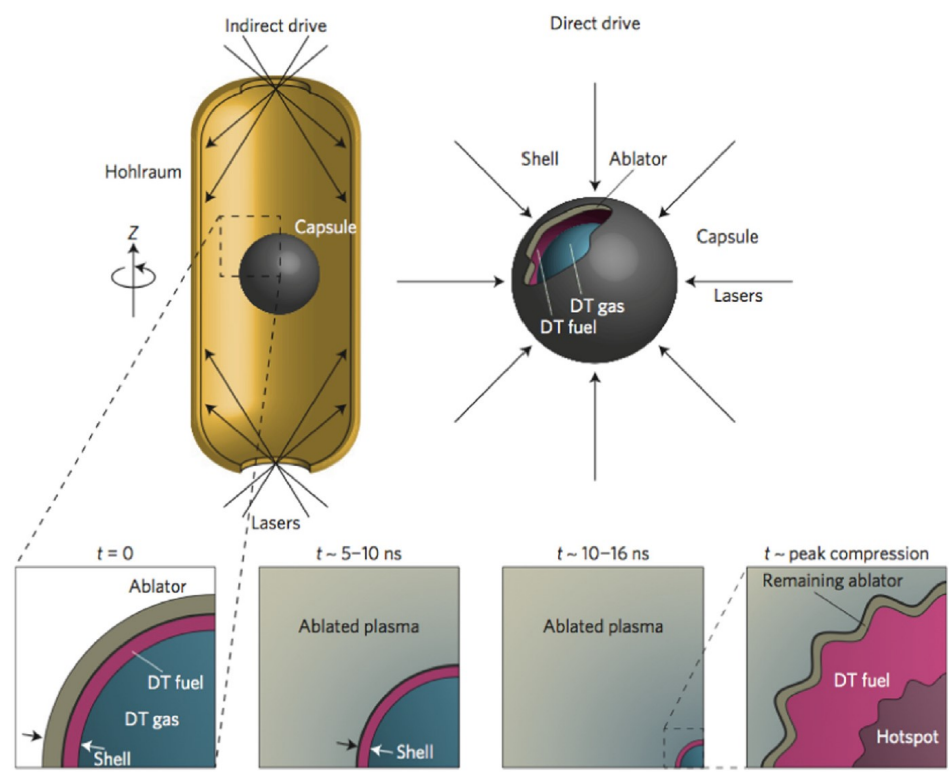

Figure 2    Laser-driven ICF is either indirectly driven (upper left) or directly driven (upper right). A spherical capsule is prepared at t = 0 with a layer of DT fuel on its inside surface. As the capsule surface absorbs energy and ablates, pressure accelerates the shell of remaining ablator and DT fuel inwards-an implosion. By the time the implosion reaches minimum radius, a hotspot of DT has formed and surrounded by colder and denser DT fuel [115].

All previous studies are about interpenetrating structures and initial perturbations of flow instabilities as some of landmark achievements in above areas and applications were published in prominent journals listed in Reference [1-115]. Very little scientific research is related to the origin of flow instabilities. The 'inside gene' of the instabilities has been ignored. Even before the instabilities start to develop, the 'inside gene' has already internally existed. It is the potential fluctuation based on the universal continuity relation. It presents as potential oscillatory wave surface in the spatial and temporal dimensions at the interface of the instabilities. Its appearance of initial perturbation could be taken as a furtive glance if the conditions were obtained [116]. In nature, the occurrence of any kind of oscillatory wave interfaces of perturbation in flow is absolutely normal because of potential fluctuations. In other words, the interfaces of instabilities have to be oscillatory in some way because of the continuity relation. After forces and energies act on the interface, initial perturbations are developed into a variety of interpenetrating structures guided by potential fluctuations and are affected by a variety of forces involved in the process.

Initial perturbations and interpenetrating structures are the structures/appearances of flow instabilities in initial stage and developed stage respectively. No scientific research has been reported to investigate the origin of flow instability. The paper presents the first attempt to conduct comprehensive analysis on origin of flow instabilities. Conducting research of flow instability without investigating potential fluctuation is like conducting research of life evolution without investigating gene. As a result, incorrect approaches and directions, irrelevant factors and conditions, contradictory statements, conflict conclusions, and errors have widely existed in previous studies. These important issues can not be adequately covered without understanding potential fluctuation, and far more significantly, major developments were simply impossible until recently, with the ability to fabricate precision initial conditions, the diagnostics tools for the experiments, and the advancement of supercomputing power. After several decades of effort, experiment, computation, and analysis have reached a level of major breakthrough. This paper brings these together, showing how observations from the experimental measurements and numerical simulations, the analytical treatments, and the phenomenological and engineering models, have been synthesized for describing the physics of the flow instabilities. The analysis would be a significant breakthrough in physics. Firstly, a new theoretical analysis of potential fluctuation based on the continuity relation is presented. Secondly, designed experimental studies of initial perturbations were conducted to provide evidences for the analysis. Finally, potential fluctuations of instabilities along the three space dimensions and one time dimension are proposed.

## Theoretical Analysis

Potential fluctuation was firstly proposed by the author at the flow interface between vapor and liquid in two-phase flow as follows [117]. Figure 3(A) presents a control volume at a point in a flow. $\rho$ is the density of the control volume including $\rho_1$ and $\rho_2$ ($\rho_1 \neq \rho_2$). The continuity relation is:

$$\frac{\partial \rho}{\partial t} + v \frac{\partial \rho}{\partial x} = 0 \quad (1a)$$

The height of interface between $\rho_1$ part and $\rho_2$ part is $H$. Both $\rho_1$ and $\rho_2$ are assumed to be constants, Eq. (1a) becomes:

$$\frac{\partial H}{\partial t} + v \frac{\partial H}{\partial x} = 0 \quad (1b)$$

Eq. (2) is valid and velocity $v$ is a constant. Along the characteristic line $\begin{cases} \frac{dx}{dt} = v, \\ t = 0: x = z, \end{cases}$ $H(t,x)$ is a constant. $H_0(z)$ is the initial value, the solution to Eq. (1b) is:

$$H(t,x) = H_0(z) = H_0(x - vt) \quad (2)$$

When a curve of $H$ exists, the time period $t$ of forming $H$ had already past, the "-" must be added into Eq. (3):

$$x = -vt \quad v \neq f(x) \quad (3)$$

Therefore:

$$H(t, -vt) = H_0(-2vt) \quad (4)$$

$H(t, x)$ can satisfy Eq. (1b) at the validation of Eq. (3). Therefore:

$$H(t,x) = A \sin\left(-\frac{k}{4v}(x-vt)^2\right) \quad (5a)$$

$$H(t,-vt) = A \sin(-kvt^2), \quad \text{when} \quad x = -vt, \quad (5b)$$

$$H(-\frac{x}{v}, x) = A \sin(-\frac{k}{v}x^2), \quad \text{when} \quad t = -x/v, \quad (5c)$$

where the initial value $H_0(z) = D \sin\left(-\frac{k}{4v}z^2\right)$ $A$, $k$, and $D$ are constants. The potential fluctuation in flows is defined in Equation (5). It has oscillatory wave behaviors along spatial dimensions and temporal dimension as shown in Figure 3(B). It is the 'inside gene' of the instabilities. By means of modeling the interface existing between two fluids and by using the boundary conditions of the differential equation, a sinusoidal type of solution has been reached. There are infinite sinusoidal solutions for the equation forming a base in which any other type of solution can be expressed. Nature mainly shows oscillatory solutions associated to lower modes because they are easier to excite since they require less energy. Potential fluctuation can not be observed in physical reality because liquid moving must require forces. After forces and energies act on a process, potential fluctuations are developed into a variety of interpenetrating structures depending on a variety of forces and energies involved in the process.

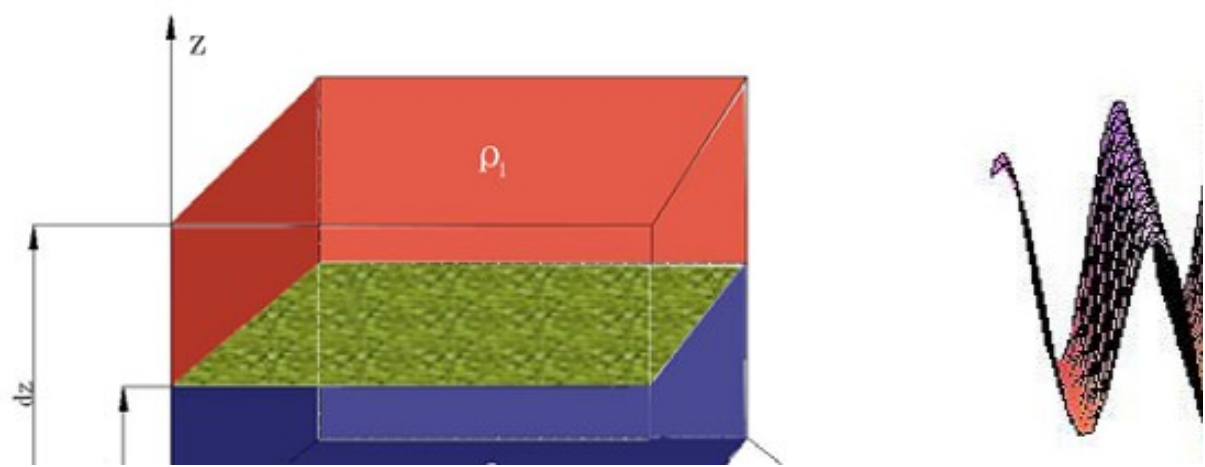

Figure 3    Wave interface between two fluids, ρ₁ and ρ₂. (A) a control volume (dx and dy with dz in depth) at a point in flow; (B) a potential fluctuation with wave behavior (*H* as function of *t* and *x*) as given in Equation (5).

The initial interface perturbation caused by potential fluctuation at the very beginning of the instabilities has wave behaviors along the space and time dimensions simultaneously. Figure 4(A) shows a three dimensional visual image of one individual initial interface perturbation with wave behavior calculated from Equation (5); *H* is a function of *x* and *t* on the interface. Similarly, *H* can be a function of *y* and *t* on the interface as well. Four dimensional initial interface perturbation is not be able to be presented in a page. A visual image of one individual initial interface perturbation in three space dimensions is shown in Figure 4(B) [118].

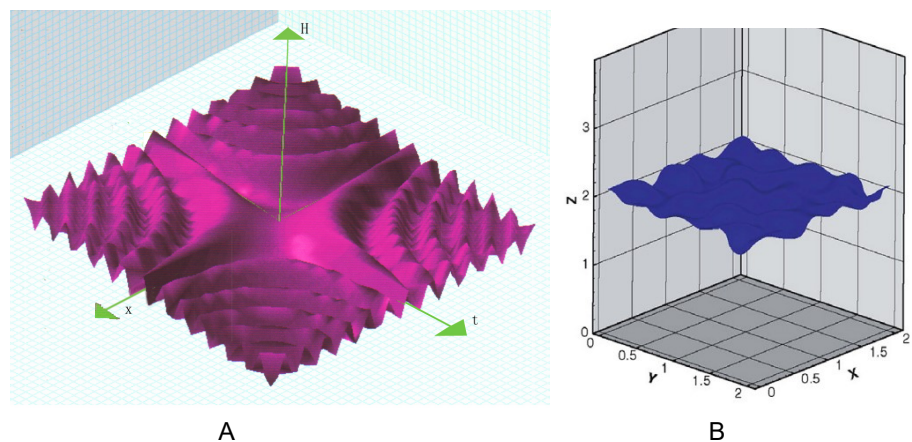

A                                              B

Figure 4    Visual image of the wave behavior of individual potential fluctuation: (A) one time dimension and two space dimensions; (B) three space dimensions [118].

The analysis of potential fluctuation is applied to perturbation phenomena of moving fluids in general flow instabilities (including but not limited to RTI, RMI, and KHI) since only the continuity relation is adopted in Equation (5). The Newton's second law of motion and the First Law of Thermodynamics are not adopted in the analysis.   To introduce the detail boundary conditions is the way of analysis based on the momentum balance and to measure accurate temperatures is the way of analysis based on the energy balance on the interface. However, to account for all the factors not only makes the situation be too complicated to be analyzed, but also it is almost an unaccomplishable mission since the experimental evidence of their effects on flow instabilities can not be directly established at present scientific testing capability [119].   Based on previous studies and to avoid their drawbacks, a unique experimental approach of fouling performances of cooling tower water in enhanced tubes is advanced to provide evidences to the analysis of potential fluctuation in the next section.

Analysis advancing new theoretical views need contain specific experimental evidences that the interpretations are distinguishable from existing knowledge. A SINE wave in visualization of the initial interface perturbation of the flow instability is the key starting assumption in the study as shown in Figure 5, where *g,* gravity; *p,* pressure; *ρ*, density; *ω,* vorticity; and circular arrows represent the velocity field [120]. Previous investigators believe that if the heavy fluid pushes the light fluid, the interface is stable; however if the light fluid pushes the heavy fluid, the interface is unstable. The interface becomes unstable with an acceleration applied in the direction of the denser fluid [121].

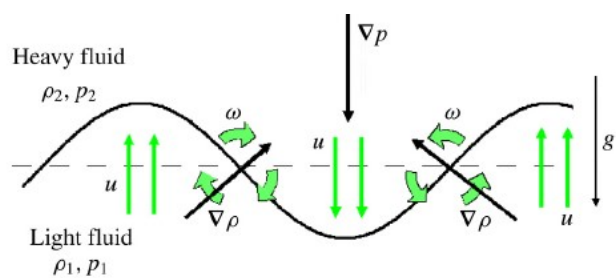

Figure 5    Assumption of Initial interface perturbation of flow instability with SINE wave [120]

Initial interface perturbation can be in the shape of SINE function as indicated in Equation (5). An example in daily life of an approximate appearance of initial interface perturbation is the scene of wind blowing over a water surface, where the wind causes the relative motion between the stratified interface of water and air as shown in Figure 6(A). The initial interface perturbations will manifest themselves through forces in the form of waves being generated on the water surface as shown in Figure 6(B), which are the result of flow instability between water and air.

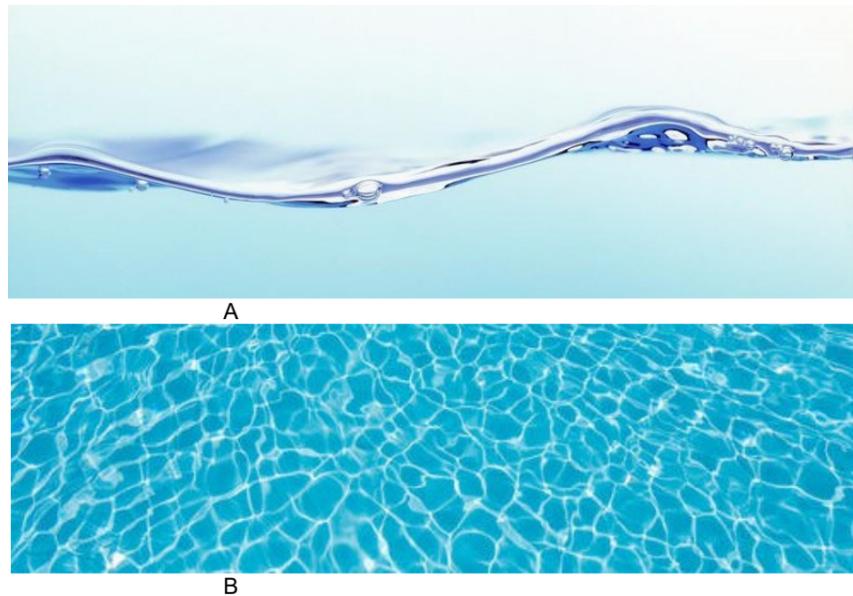

Figure 6     Flow instability by a wind blowing over water surface:    (A) one initial interface perturbation; (B) pattern created by initial interface perturbations.

## Experimental evidences of potential fluctuation in flow instabilities

**Molecular scale experiments**

It is a challenge to conduct experiments on initial interface perturbation at molecular scale since small irregularities will inevitably grow and break the repetitive symmetry. However, fouling deposit layer from cooling tower water provides figures of initial interface perturbations to obtain valuable information as stated in the author's previous paper [122]. Evaporative cooling that takes place in a cooling tower increases the overall percentage of non-evaporative impurities in the water. Hardness in cooling tower water consists mainly of dissolved salts ($CaCO_3$ and $MgCO_3$); when this solution contacts the tube wall (heated surface), crystal deposits begin to form. The solubility of $CaCO_3$ and $MgCO_3$ decreases with increasing temperature; these salts along with rust and dust particles will precipitate on the wall of the heated condenser tubes forming a fouling layer. Deposit were removed from the evaluated tubes at the end of one cooling season; the deposit was then analyzed for chemical composition and showed that 61% of the deposit is calcium carbonate and the remaining 39% is characterized as non-crystallizing particles [silica, iron phosphate/silicate (from rust and dust) and copper oxide (from the copper tubing)]. The fouling deposits firmly attach to the surface. Rust and dirt particles are present in cooling tower water as a result of particulate being deposited in the water from the cooling air. Suspended particle (salt, rust, dust, etc.) size in the cooling tower water is approximately 3.0 μm. Tube fouling occurs when the particulate in the cooling water deposits on internal wall of the tube.   Transport of particles to the tube wall produce deposits on the surface; this is the result of the diffusion particle transport of the 3.0 μm particles in the cooling tower water. The particles move with the fluid and are carried to the wall by Brownian motion at molecular scale.

The density of the fouling layer $\rho_2$ (fouling deposits and water) is larger than the density of the fluid $\rho_1$ (water only) above the fouling layer. The fouling layer and the fluid provide the density difference and velocity difference that are required to observe flow instability on the surface of the fouling layer.   In a newly built eight-story classroom building on campus, a series of experiments were performed in order to evaluate heat transfer performance of enhanced tubes under fouling conditions using cooling tower water. Testing took place in a test evaluation shell-and-tube condenser that was installed in parallel to the main condenser of 866 kW water chiller. The evaluation condenser unit contained four helical tubes that were 3.7m long. Tubes were installed horizontally as pairs for each tube (Tube 1 and Tube 2 in Table 1). The test unit used the same water from the cooling tower as the main condenser. Both tubes evaluated were identical for outside enhanced surface with an outer diameter of 15.54 mm, tube wall thickness of 0.5 mm, included angle between sides of outside ribs = 41°, outside fin tip thickness = 0.024 mm, with an identical outer surface (1024 fins that had a height of 0.90 mm).   Tested sample tubes are shown in Figure 7. As can be seen they are very different and the geometric parameters for inside enhanced surface of the two tubes are summarized in Table 1. Tube pairs are used in order to determine transient performance differences; one tube is not fouled (maintained clean by regular manual cleaning) and the other tube is never cleaned and fouling occurs over a period of time. One pair of Tube 1 and another pair of Tube 2 were tested at same time. Thermal resistance of the fouling layer ($R_f$) is determined by evaluating the difference in the total heat transfer coefficient between the fouled tube and clean tube. Cooling tower water (composition: calcium hardness - 651ppm; Resistance: 1590 - 1810 μohm with a pH of 8.52) was circulated through the test condenser unit with mass flow rate of G = 740 kg/($m^2$s) using a single pass.

Inlet cooling tower water temperature was 29.3°C, Outlet water temperature was 32.8 °C, with a refrigerant condensation temperature of 35.7 °C. The fouling data were taken every 48 hours, over 2580 hours of operation (one cooling season). Testing was conducted a second time during the following cooling season in order to evaluate the repeatability of the data.

Table 1    Internal geometric parameters of the enhanced condenser tubes that were evaluated

| Tube | number of rib starts, $n_s$ | internal rib height, $e$ (mm) | helix angle, $\alpha$ (degree) | axial element pitch/ internal rib height, $p/e$ |
|---|---|---|---|---|
| 1 | 10 | 0.43 | 35 | 9.88 |
| 2 | 45 | 0.33 | 45 | 2.81 |

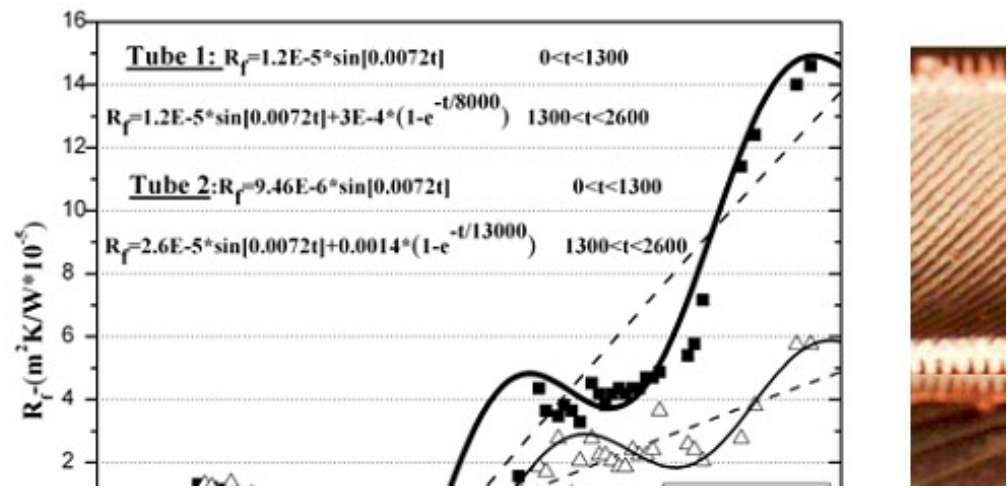

Figure 7    Results from the transient tube fouling evaluation using recirculated cooling tower water. Data were taken every 48 hours over a total time period of more than 2580 hours (cooling season length). Copper tube outside diameter is 15.54 mm, made with 1024 fins/in (height - 0.90 mm) on the outer surface; inside tube parameters of the helically ridged tubes were: number of rib starts (10 for Tube 1, 45 for Tube 2), helix angle (25-deg for Tube 1. 45-deg for Tube 2), and height (0.43 mm for Tube 1, 0.33 mm for Tube 2). (A) Fouling data with curve fits for Tube 1 and Tube 2. (B) Photographs of the tubes evaluated (Tube 1 and Tube 2) at the end of one cooling season.

Figure 7 shows the experimental results of the long-term fouling evaluation that was performed in the enhanced tubes, Tube 1 and Tube 2 (two different internal helically ridged tubes). Since the experiment can only be performed during the operating hours of a cooling season, the data that details the full development stage is impossible to obtain. After an introduction period, the fouling of the cooling tower water demonstrates asymptotic behavior $B(1-e^{-t/b})$, with $R_f$ is proportional to $H$ (thickness of the fouling layer). Since $A_t \sin(\lambda_t t)$ is more suitable than $A_t \sin(\lambda_t t^2)$ for correlating data points, $A_t \sin(\lambda_t t)$ is used in the development of a correlation. Figure 7 (A) shows wave curves of the two evaluated tubes for both the introduction period (0-1300 hours) and the development period (1300 - 2580 hours). It shows that the oscillatory asymptotic fouling curves (solid lines) correlated well with the experimental data; The dotted lines are the asymptotic fouling curves.   The oscillatory asymptotic fouling curves are defined using the general format, $R_f = A_t \sin(\lambda_t t) + B(1-e^{-t/b})$. Equations (6 a, b) detail the oscillatory asymptotic fouling curves for Tube 1 and equations (7 a, b) define the curves for Tube 2.

$$R_f = 1.2 \times 10^{-5} \times \sin(0.0072t) \qquad 0 < t \leq 1300 \qquad (6\text{ a})$$

$$R_f = 3.0 \times 10^{-4} \times \left(1.0 - e^{-t/8000}\right) + 1.2 \times 10^{-5} \times \sin(0.0072t) \quad t \geq 1300 \qquad (6\text{ b})$$

$$R_f = 9.46 \times 10^{-6} \times \sin(0.0072t) \qquad 0 < t \leq 1300 \qquad (7\text{ a})$$

$$R_f = 1.4 \times 10^{-3} \times \left(1.0 - e^{-t/13000}\right) + 2.61 \times 10^{-5} \times \sin(0.0072t) \quad t \geq 1300 \qquad (7\text{b})$$

As can be seen in equations (6) and (7), the temporal fluctuation cycle ($\lambda_t$) for the introduction and the development periods are the same ($\lambda_t$=0.072) for both evaluated tubes; it indicates that $\lambda_t$ is not a function of the geometric parameters of the surface. The deposit layer temporal fluctuation amplitude ($A_t$) for the development period is larger than that the introduction period in both tubes; this is a result of different thickness of fouling deposits occurring in the development period and result in different overall values of $A_t$ for Tube 1 and Tube 2. These differences are the result of different stress conditions in the two tubes, including: (i) flow being interrupted by the rib; (ii) boundary layer close to the wall is separated and then reattaches to the wall downstream of the rib (at an approximate distance of 5 to 8 times the rib height); (iii) reattachment point vanishes when the pitch of the rib is reduced to less than about five times of rib height, forcing the main flow to "glide over" the ribs and produce secondary flows between the ribs. Rib pitch to height ratio ($p/e$) is greater than 5 for Tube 1 ($p/e$≈9.88), for this case the fluid effects the fouling deposit near the ribs by producing a larger wall shear and disperses the fouling deposit accumulation. When $p/e$ <5 (as the case of Tube 2 with $p/e$≈2.81) there is less effect of the fluid on the fouling layer; this results in a smaller wall shear that produces sheltered low velocity areas that allow the initial growth of the fouling deposit to take place (producing the growth of the fouling deposit). Additionally, $A_t$ in Tube 2 is larger than that in Tube 1, resulting from thicker fouling deposits in Tube 1; The $A_t$ difference is caused by difference of wall shear stress between Tube1 and Tube 2. Photographs of Tube 1 and Tube 2 in the fouled condition (at the end of 2580 operating hours) are shown in Figure 7(B) with the corresponding data given in Figure 7(A). Figure 7(B) shows the wave appearance of the fouling deposit along the center line. Eq. (5c) can be simplified to be $H=A_x \sin(\lambda_x x)$, with the spatial fluctuation cycle ($\lambda_x$) being the same in the two tubes as showing in Figure 7(B). As state in previous paragraph, Figure 7 can be used to identify the initial interface perturbations. For an initial interface perturbation, $\lambda_x$ and $\lambda_t$ are related to density and velocity, they are not related to geometric parameters; $A_t$ and $A_x$ are related to density and velocity and are effected by the stresses which are related to geometric parameters.

Temporal wave behavior is shown in Figure 7(A); spatial wave behavior is seen in Figure 7(B). Figure 7 details the experimental data for the initial interface perturbation caused by the potential fluctuation as shown in Figure 4. Some studies of wave asymptotic fouling curves in literature include: (i) fouling data for an extended surface heat exchanger and a mass accumulation probe used in a diesel exhaust environment [123]; (ii) corrosion fouling [124]; (iii) magnetite particulate deposition from water flowing in aluminum tubes using an X-ray technique [125] in Figure 8(A); (iv) cooling tower water fouling performance of two kinds of brazed-plate heat exchanger (BPHE) [126] in Figure 8(B).

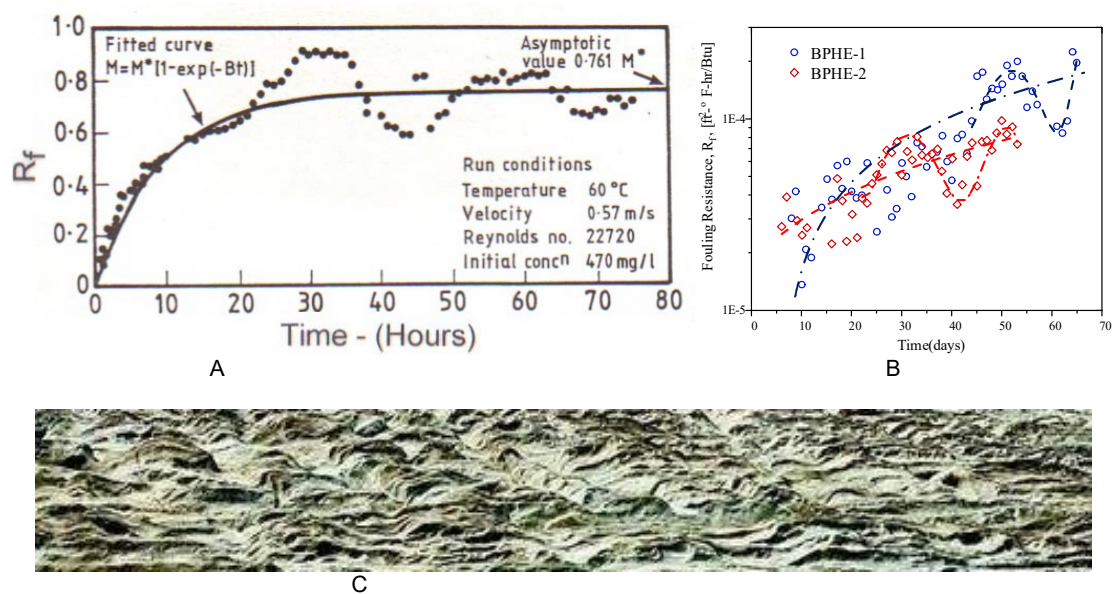

Figure 8    Wave behavior of a fouling deposit: (A) magnetite particulate deposition for water flowing in aluminum tubes [125]; (B) cooling tower water fouling deposit in a brazed-plate heat exchanger (BPHE) condenser [126]; (C) water flow effect on a river bed.

As can be seen in Figure 8(B), the wave asymptotic fouling curves for the fouling resistance of cooling tower water in two different plate heat exchangers (BPHE-1 and BPHE-2, each with a different plate surface) have the same values for $\lambda_t$ and $A_t$ as they are not functions of the geometric parameters. Temporal wave fouling, as examples shown Figure 7(A) and Figure 8(A, B), has been reported in literature for more than fifties years, however it has not yet been theoretically analyzed. Instead researchers have focused on the asymptotic fouling and have ignored temporal wave fouling since they thought that its wave behavior was generated as a result of experimental uncertainty. Previous researchers have usually reported their fouling data using temporal methods as shown Figures 8(A, B); typically not considering spatial methods. Some examples of spatial wave deposit layers can found in nature. One example is water flow effects on the river bed as shown in Figure 8(C), here the effect on the surface structure is produced from small sand particles, water flowing a river bed produces the density and velocity difference in the flow. It is interesting to observe that the surface structure in Figure 8(C) is similar to the wave pattern in Figure 6(B).

The deposition process that forms the fouling layer in Fig.7 is similar to the digital process that a 3D printer constructs objects, being slowly built layer-by-layer by 3.0 µm particles in the cooling tower water. Equation (5) describes the 'digital model', with the fouling layer being the 'printing layer'. This can be used to identify the initial interface perturbations. It is the uniqueness of the fouling experiment at molecular scale. In general, the initial interface perturbation varies in space and time, $A(x,y,z,t)$ and $\lambda(x,y,z,t)$. $\lambda(x,y,z,t)$ and $A(x,y,z,t)$ are determined by densities ($\rho_1$ and $\rho_2$) and velocities ($v_1$ and $v_2$) of the two fluids involved in the flow instability.

**Micro scale experiments in µm**

In recent decades, Silicon Carbide and Gallium Nitride power electronic devices have been extensively studied and employed in various industrial applications because of their excellent high power and high frequency performance. Fig. 9 presents the cross-sectional SEM images of the chip, and the geometrical parameters are listed in Table 2.

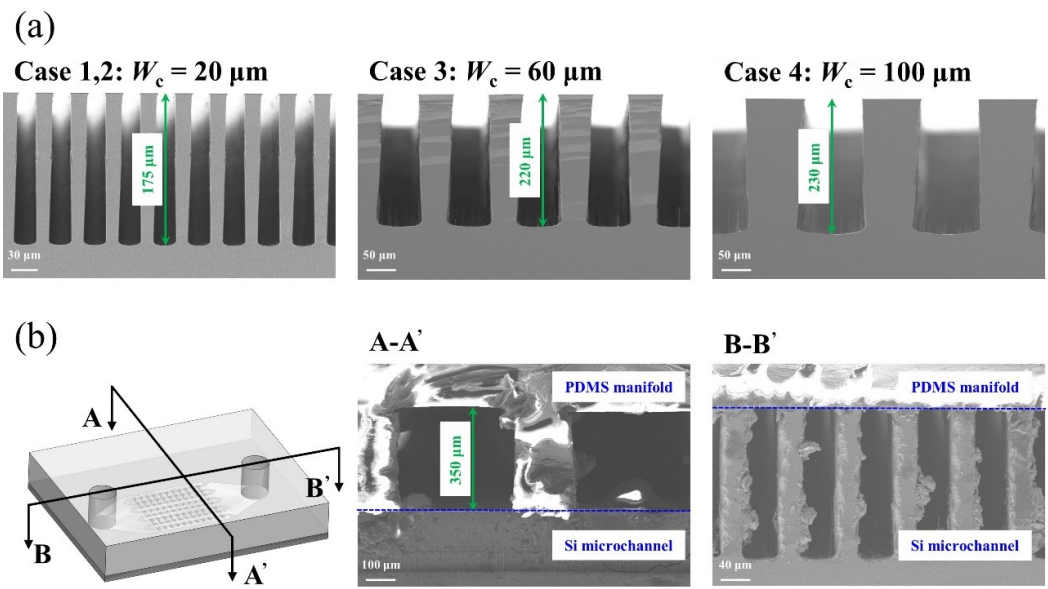

Figure 9  Typical fabrication results of chip. (a) Cross-sectional SEM image of the microchannel in dummy chips. (b) Cross-sectional SEM image of the dissected chip.

Table 2  Geometric size parameters in cooling chips

| Substrate | Parameter | Symbol | CASE 1 | CASE 2 | CASE 3 | CASE 4 |
|---|---|---|---|---|---|---|
| Chip | Length | $L$ | 17.5 mm | | | |
| | Width | $W$ | 14 mm | | | |
| | Area of heater | $A_{heat}$ | 5×5 mm$^2$ | | | |
| Microchannel | Height of Microchannel | $H_c$ | 175 μm | 175 μm | 220 μm | 230 μm |
| | Width of Microchannel | $W_c$ | 20 μm | 20 μm | 60 μm | 100 μm |
| | Fin width of Microchannel | $W_{c,f}$ | 20 μm | 20 μm | 60 μm | 100 μm |
| | Hydraulic diameter of Microchannel | $D_h$ | 36 μm | 36 μm | 94 μm | 139 μm |
| | Number of Microchannels | $N_c$ | 125 | 125 | 42 | 25 |

Fig. 10 shows the schematic diagram of the experimental setup and test facility.   In the experimental loop, the coolant flows in a closed loop through a system pressure regulator, a 50 μm filter, a gear pump, a Coriolis mass flow meter, a heat exchanger with a thermostatic water bath, a 2 μm filter, a preheater, a test section and a cooler successively. A system pressure regulator is attached between the coolant injection pipeline and the experimental loop. Two filters are used to capture particles larger than 2 μm to prevent blockage of the channels. The gear pump can supply a maximum flow rate of 1.0 litre per minute and a maximum pressure difference of 100 psi. The Coriolis mass flow meter can measure the mass flow rate of the coolant flow. The stainless steel preheater located upstream of the test section can raise the inlet coolant temperature to a required value, using a manually controlled DC power. Two heat exchangers are located downstream of the test section to cool the coolant down to a specified temperature. The test module is the core of the experimental setup, consisting of a movable sample stage, a PEEK support plate, a PCB board and the cooling chip. The cooling chip is soldered on a PCB board for the convenience of applying Joule heating and determining the location. The heater pad is soldered onto the PCB, which is then fastened to a PEEK support plate using bolts. As showed in Fig. 11, a high-speed microscope camera is positioned above the PDMS manifold plate to visualize fluid flow. The transparency of the PDMS manifold plate is sufficient for observing the fluid flow state during the heat transfer process.

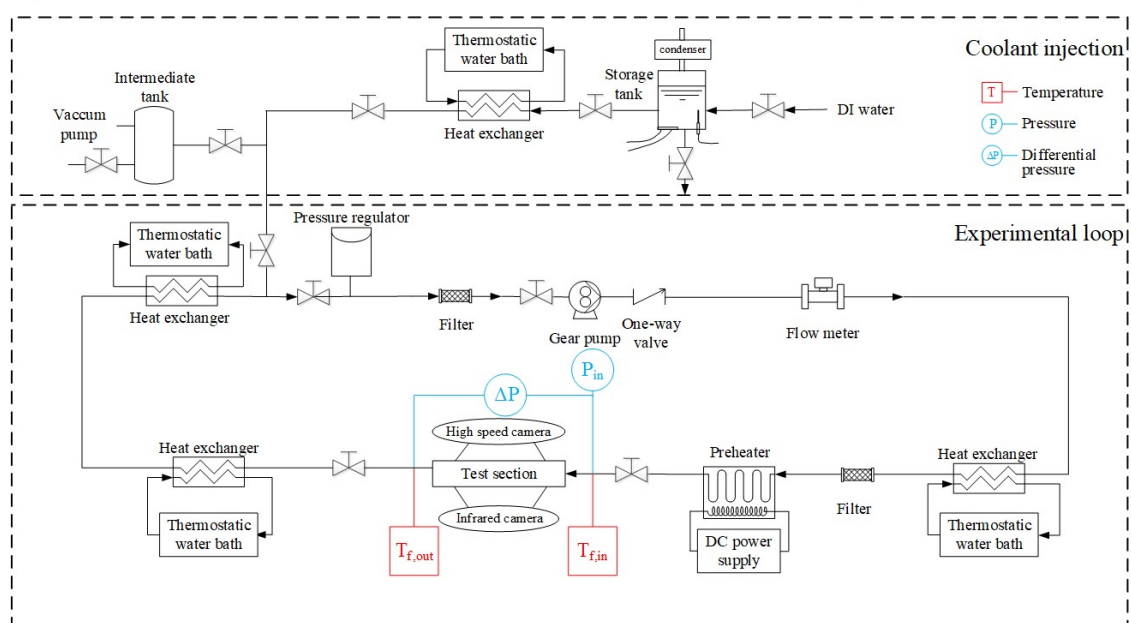

Fig. 10.   Schematic diagram of the MMC heat sink experimental setup and test facility.

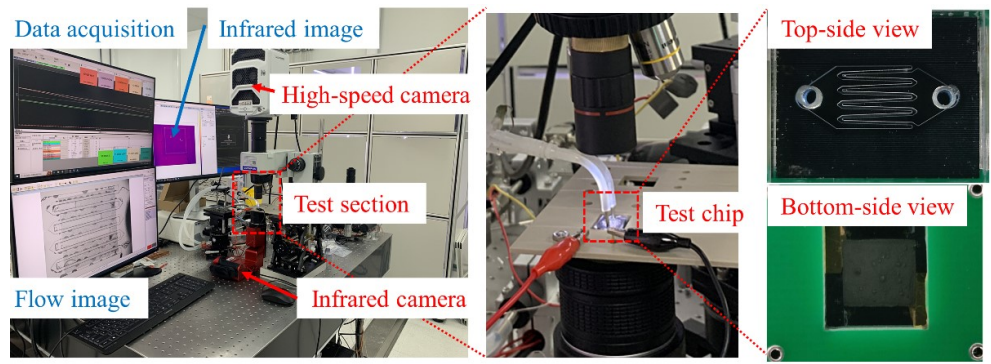

Fig. 11. Physical maps of (a) the experimental setup and test section; (b) the assembled test chip; (c) the test chip and a custom PCB board.

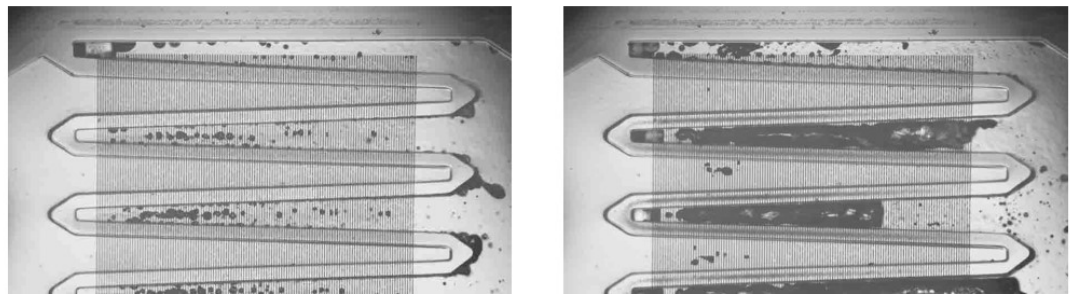

Fig. 12  Comparison of typical heater surface boiling flow regimes using water at flow rate of 4 g/s in Case 1 for: (a) $q_{eff}$ = 672 W/cm$^2$; (b) $q_{eff}$ = 838 W/cm$^2$; (c) $q_{eff}$ = 1018 W/cm$^2$.

The visualization of the flow regime in Fig. 12 will show interface perturbation at micro scale of μm by Image binarization processing as shown in Fig. 17.

**Conventional scale experiments in mm**

As shown in Fig. 13, the experimental apparatus consists mainly of a refrigerant circuit and a deionized-water circuit. The refrigerant is delivered from the tank to the preheater by a gear pump. A Coriolis flowmeter is used to measure the mass flowrate of the refrigerant. A water bath and the preheater are used to establish the thermodynamic state of the refrigerant that enters the test section. The test tube is jacketed by a deionized water annulus. The temperature and the flowrate of the deionized water is controlled by a bath, which in turn, controls the heat flux to the test section and the exit quality of the refrigerant.  Temperature and pressure sensors were installed at the entrance and the exit of the test section for monitoring the states the of refrigerant and deionized water. The refrigerant flow exiting the test section, is condensed by an oil-cooled condenser and deliver as liquid to the storage tank to complete the cycle. Temperatures, pressures and flowrates were recorded by the data collector and saved to the computer, with a recording interval of 20 seconds. The exterior surfaces of the preheat section, the test section and the subcooling were insulated with a cotton insulation.  In addition, polyurethane foam was applied to the outside of the casing of the experimental section before each experiment insulate the experiment from the environment.

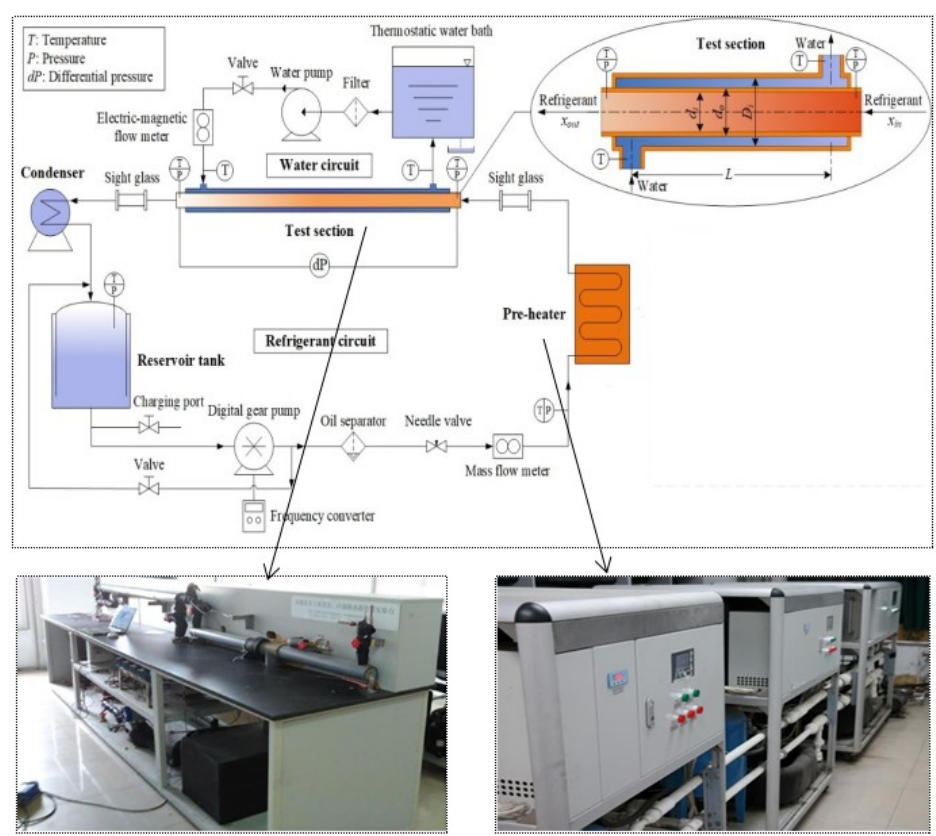

Fig.13    Schematic diagram and front and rear views of the experimental equipment

The test section is manufactured as a tube-in-tube heat exchange fitted with the test tube as the inner tube. The effective heat transfer length of the test section is 1.8 m, and a copper tube with an outer diameter ($d_i$) of 17.0 mm is selected as external tube. During two-phase flow experiments, refrigerant flows inside the test tube while heating or cooling water flows through the annulus. Main geometric parameters for tested smooth and EHT tubes are listed in Table 3.

Table 3    Geometric parameters of two tested tubes.

| Parameter | Smooth tube | EHT tube |
|---|---|---|
| Material | Cu | Cu |
| Outer diameter, $d_o$ (mm) | 12.70 | 12.70 |
| Inner diameter, $d_i$ (mm) | 11.30 | 11.28 |
| Wall thickness, $\delta$ (mm) | 0.70 | 0.71 |
| Dimpled/protrusible diameter, $d_p$ (mm) | - | 4.40 |
| Height of dimples, $h$ (mm) | - | 1.71 |
| Pitch of dimples, $p$ (mm) | - | 9.86 |
| Number of dimple arrays, $N_{array}$ | - | 4 |

The enhancement of the EHT tube can be divided into two categories: background arrays of petal-shaped dimples and regular dimple matrixes as presented in Fig. 14 (A to D).

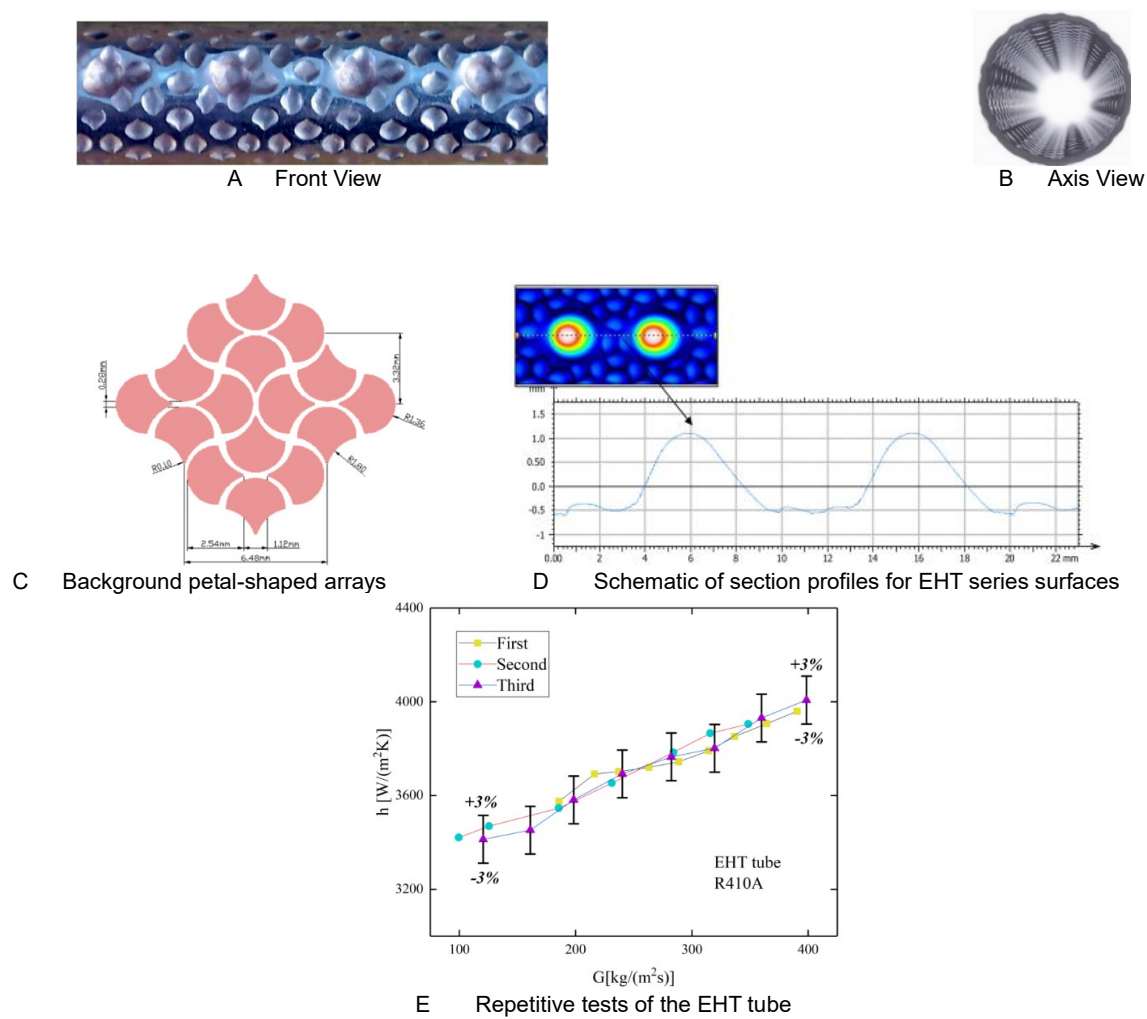

A    Front View                                    B    Axis View

C    Background petal-shaped arrays      D    Schematic of section profiles for EHT series surfaces

E    Repetitive tests of the EHT tube

Figure 14    (A) and (B) Structures of EHT tube; (C) Geometrical parameters of the background petal-shaped arrays; (D) Schematic of section profiles for EHT series surfaces (Unit: mm); (E)   Repetitive tests of the EHT tube

To validate the accuracy of the experimental results, the condensation flow repetitive tests of the EHT tube were conducted. The experimental results are repeatable within a random error band of 3%, which shows that the experimental data have good reproducibility.

Flow patterns observed in co-current, two-phase flows in horizontal tubular channels are complicated by the asymmetry of the phases, which are influenced by the inertia stresses.   To observe the two-phase flow pattern in a tube, researchers usually install the sight glass at the end of the test section. The sight glass is generally made of quartz glass with high pressure-bearing capacity, with a length ranging from 100 mm to 120 mm. The flow pattern at the exit of the test tube can be observed through the sight glass. In this study, a see-through section with an inner diameter of 11 mm is installed at the end of the test section to observe flow patterns in smooth and EHT tubes. As illustrated in Fig. 15, a 10-mm-long silica glass tube used as a sight glass for flow visualization is located on center axis position between two stainless steel flanges. A 1.5-mm-thick nylon gasket is placed between the silica glass tube and flange to prevent collision between the sight glass and stainless-steel flange and ensure air-tightness of this device. Extending from the flange on two sides is a stainless steel hexagonal inner wire joint welded to the flange, a reducing hex nipple, and a hollow nut. In addition, two 11.3-mm-ID copper tubes are welded seamlessly to the main body of the sight glass and test tube or exit section of the test section. To reduce environmental heating, metal parts are insulated with polyurethane foam insulation material wrapped by a PVC plastic shell. The main body of the sight glass is wrapped by 6-mm-thick rubber insulation cotton.

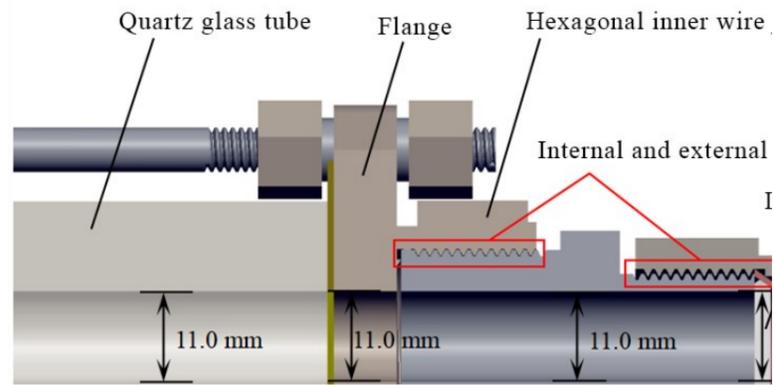

Figure 15    Broken-out section view of the acquisition equipment for in-tube flow patterns.

The characteristic parameters of vapor-liquid two-phase flow distribution, such as average liquid level height and cross-sectional void fraction, are mainly affected by the vapor quality, mass flow rate, thermophysical properties of the working medium, and wall shear stress. As long as the vapor quality, mass flow rate, and thermophysical properties of the working medium remain unchanged in the tested tube and in the sight glass, the flow pattern in the tested tube and in the sight glass is the same. If the inner wall of the sight glass is smooth and clean, and the inner diameter is close to that of the test tube, then the difference of the flow pattern between the interior of the glass tube and the outlet of the test tube can be neglected. The inner diameter of the smooth tube and EHT tube is 11.30 mm and 11.28 mm, respectively. The inner diameter of the copper tube welded with the test tube is 11.30 mm. However, due to the processing difficulty and the need to fit the glass tube with its small diameter and large wall thickness precisely, the quartz glass tube molded by the blow-molding process has an inner diameter of 11.0 mm. The inner diameter of the test tube and that of the glass tube deviate by 0.28 to 0.3 mm. To reduce the influence of the sudden shrinkage of the flow channel on the flow pattern in the pipe, the inner diameter of the leak-proof clasp matched with the copper pipe is 11.2 mm. In addition, the inner diameters of the straight reducing pair wire, flange, and nylon washer are each 11.0 mm. Therefore, possible turbulence resulting from a sudden enlargement or contraction is avoided in this acquisition equipment. After experimental conditions reach a stable state, a Phantom high-speed camera (VEO-1010) is utilized to distinguish flow regimes with an image acquisition frequency of 4000 fps and a resolution of 1280 × 960 pixels.

Fig. 16 presents evaporation and condensation flow regimes using R410A in smooth and EHT tubes. Three samples of typical experimental observation with scale bars are shown in Fig.16 (D). As shown in the following pictures taken by the high-speed camera, all the pictures have the same view with a height of 11.3 mm, and a fixed aspect ratio of 3. Additionally, the refrigerant flows from left of the view to right. The two-phase flow interface is relatively smooth at low vapor qualities, as vapor quality and mass flow rates increase, the liquid-vapor interface no longer has a stable, smooth shape due to inertia and surface tension stresses. For the same mass flow rate, the relative velocity of vapor phase increases as vapor quality increases; Friction between two phases causes fluctuations in the liquid-vapor surface to increase, becoming more chaotic and eventually changing into annular flow at higher vapor qualities and mass flow rates.

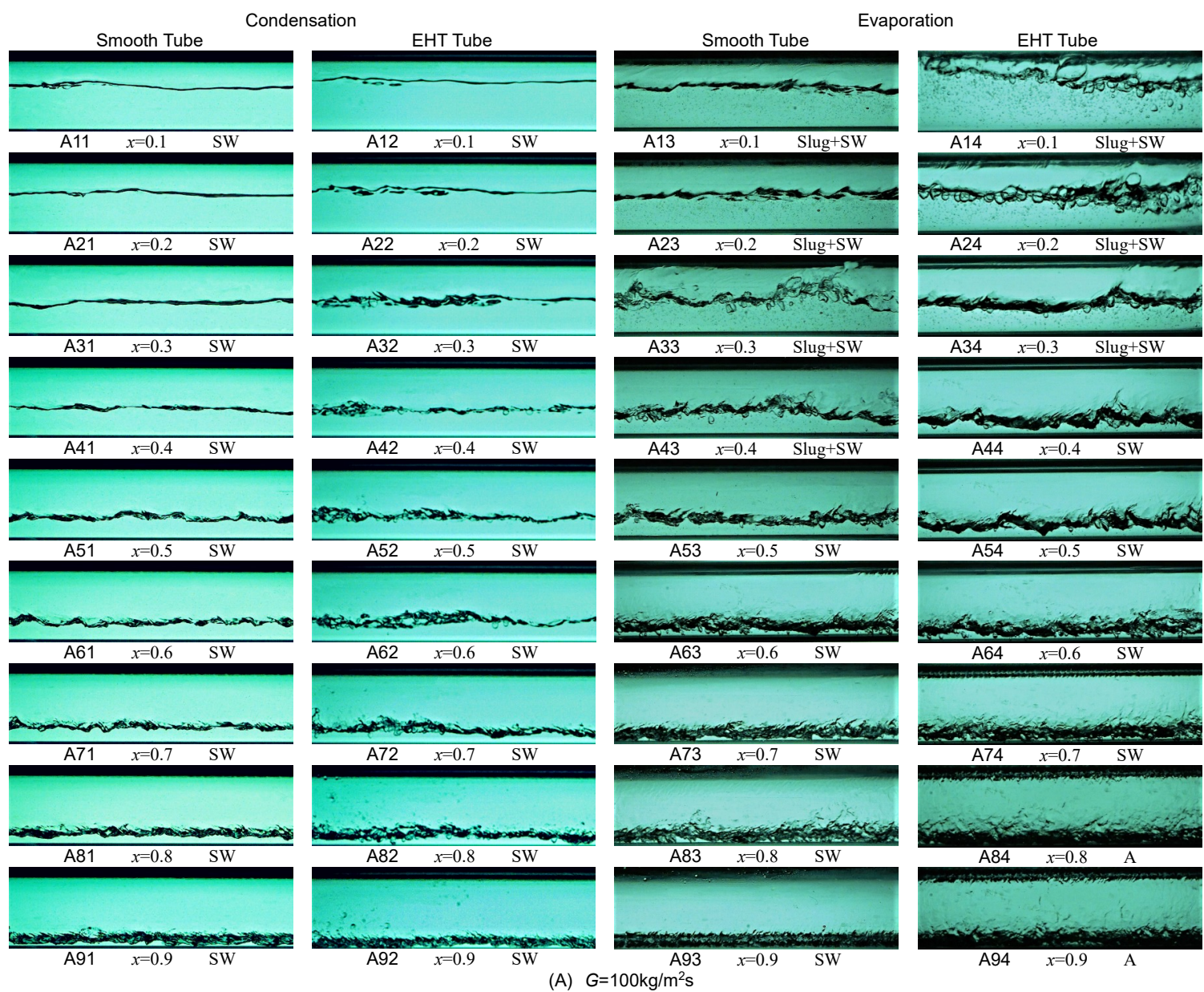

(A)  $G=100 \text{kg/m}^2\text{s}$

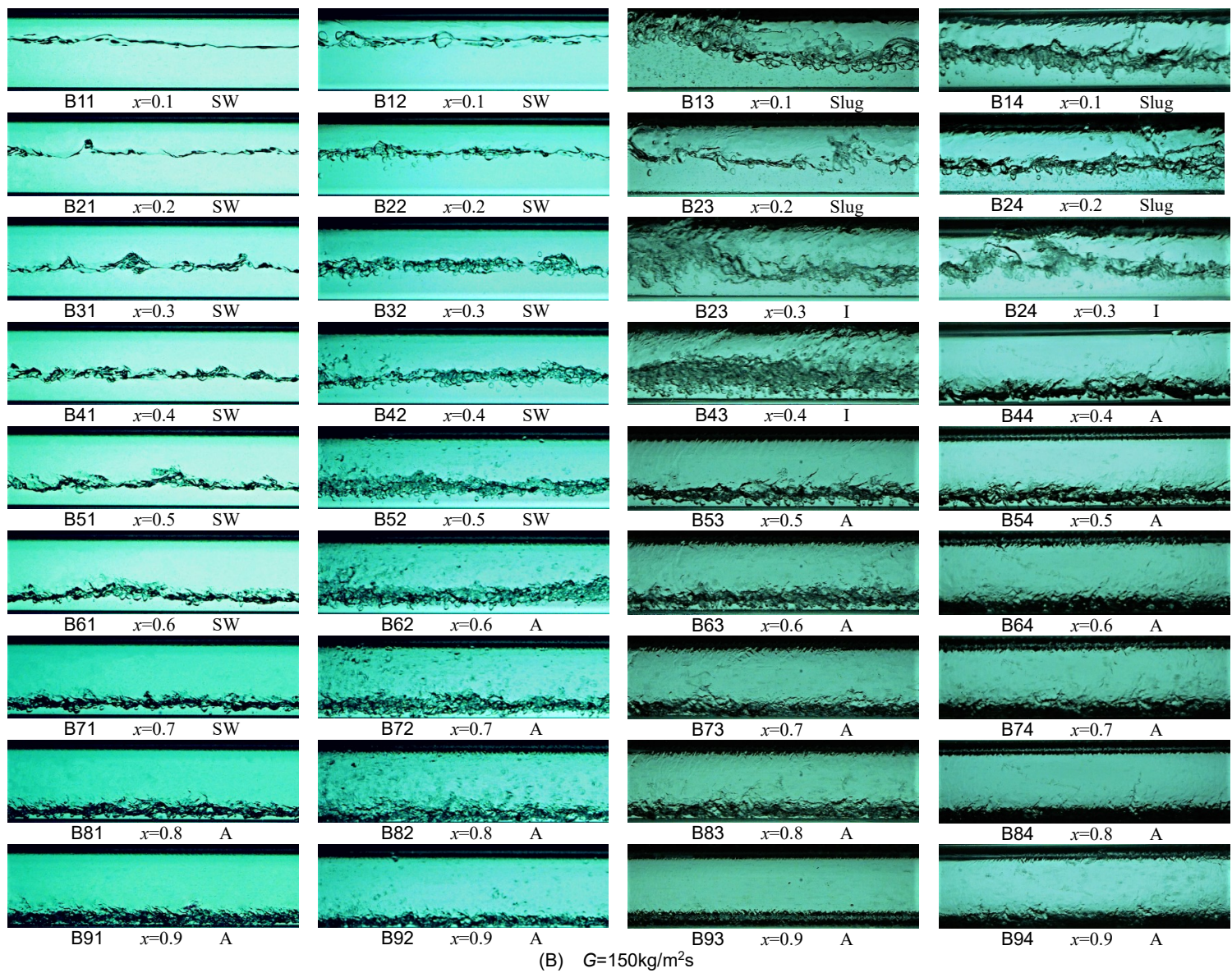

(B)  $G$=150kg/m²s

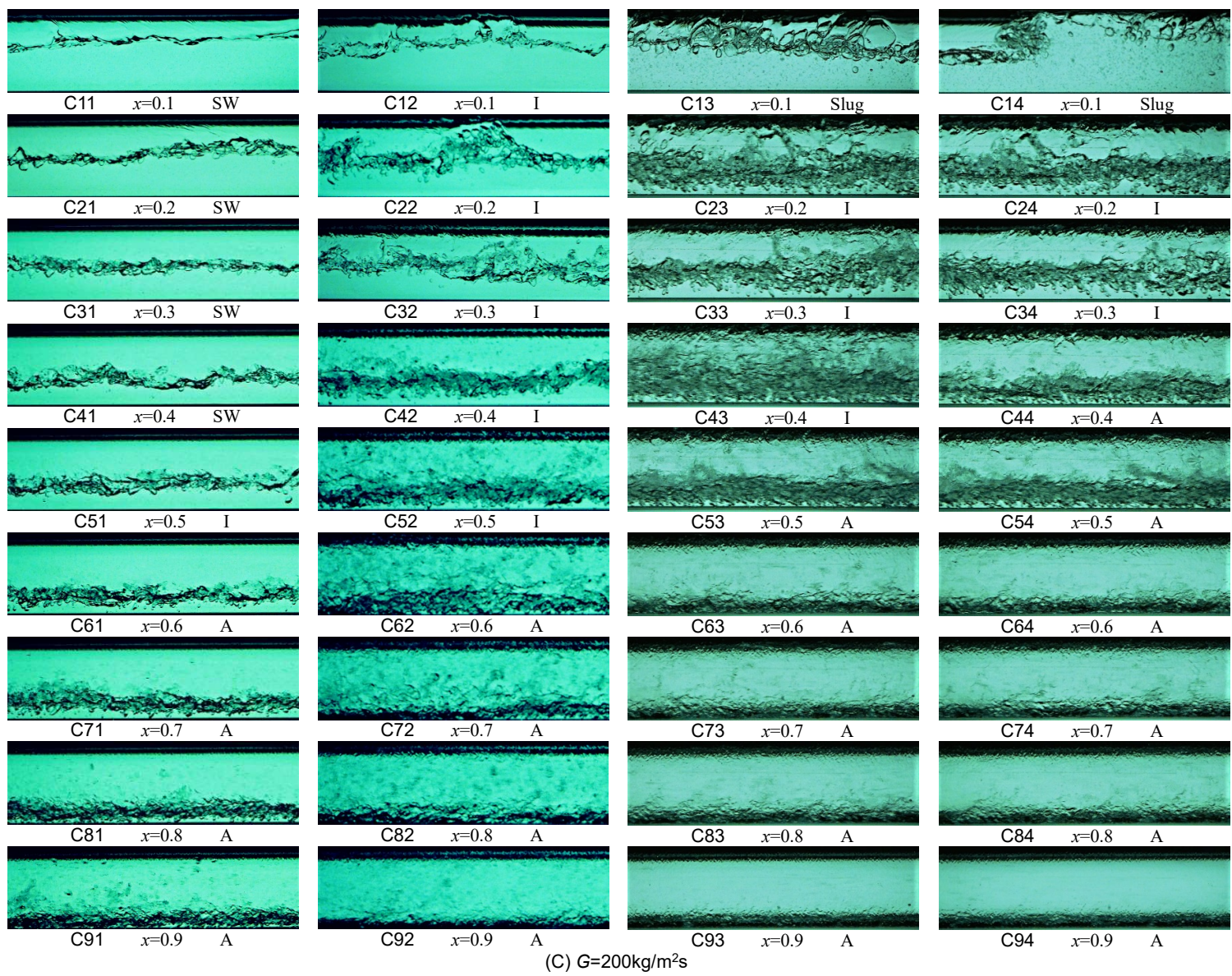

(C)  $G$=200kg/m²s

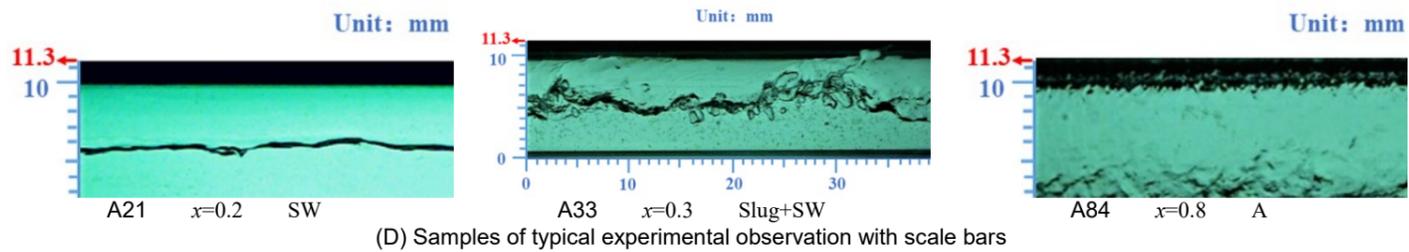

(D) Samples of typical experimental observation with scale bars

SW: stratified-wavy flow; Slug: Slug flow; I: intermittent flow; A: annular flow.

Figure 16  Evaporation and condensation flow regimes using R410A in smooth and EHT tubes: (A)$G$=100kg/m$^2$s; (B)$G$=150kg/m$^2$s; (C)$G$=200kg/m$^2$s.  (D) Samples of typical experimental observation with scale bars.

A sample of pictures of flow regimes in Fig. 16, Fig. 17 (A) can be presented as image in Fig. 17 (B) after binarization processing. Liquid-vapor interface level contour can be extracted in Fig. 17 (D), which is a wave curve.

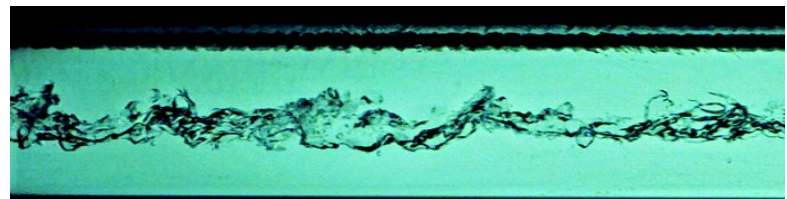

(A)    A sample of typical experimental observation

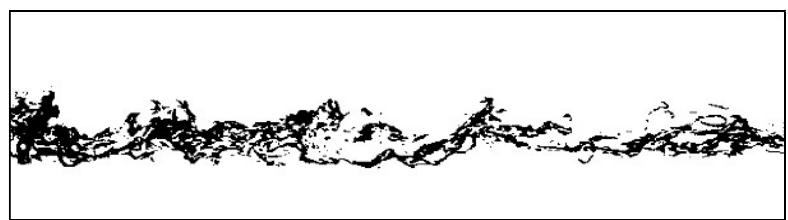

(B)    Image binarization processing

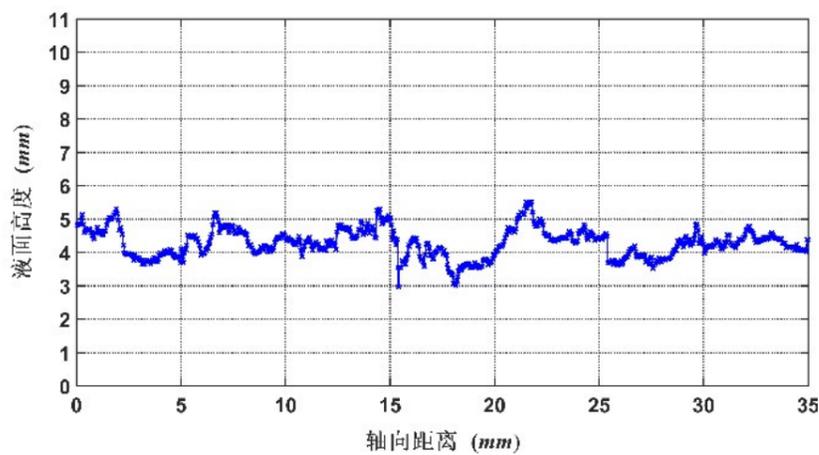

(C)    Liquid-vapor interface level contour

Figure 17  Liquid-vapor interface wave curve in two-phase flow: (A) A sample of typical experimental observation; (B) Image binarization processing: (C) Liquid-vapor interface level contour

Similarly, liquid-vapor interface level contour, a wave curve can be extracted from typical heater surface boiling flow regimes in Fig. 12 as well. In other words, liquid-vapor interface wave curve in Figure 17 (C) presents experimental evidences of interface wave curve in two-phase flow at conventional scale and micro scale. The interface wave curve in two-phase flow, which has oscillatory wave behaviors along spatial dimensions, is the development of potential fluctuation as defined in Equation (5).

## Conclusions

Potential fluctuation presents potential SINE wave surface in the three spatial and one temporal dimensions at the interface between two moving fluids. Potential fluctuations exist at instability interfaces between two fluids whenever one fluid is accelerated by the other one (including but not limited to RTI and KHI). They are the origins of flow instabilities. Even before the flow instabilities begin to develop, potential fluctuation has already internally existed in flow. It acts as the 'inside gene' to control the development of flow instabilities. The theoretical analysis is conducted based on two fundamental principles: the continuity relation and the relation among distance, velocity, and time. Potential fluctuation as defined by Equation (5) may be a new physics concept for universal flow instabilities of molecular, micro, macro, and cosmic phenomena.

Initial interface perturbation can be described by SINE function of amplitude $A(x,y,z,t)$ and wavelength $\lambda(x,y,z,t)$ based on the theoretical analysis of potential fluctuation and confirmed by fouling performance of cooling tower water in enhanced tubes. $\lambda(x,y,z,t)$ and $A(x,y,z,t)$ are determined based on the densities ($\rho_1$ and $\rho_2$) and velocities ($v_1$ and $v_2$) of the two fluids involved in the flow instability. $A(x,y,z,t)$ is affected and $\lambda(x,y,z,t)$ is not affected by the stresses on the wave interface. More direct experimental evidence of potential fluctuation, with the advancement of the diagnostic tools developed for the experiments and supercomputing power, are needed to obtain specific information of Equation (5).

Previous researchers have adopted SINE wave interface as the key starting assumption to develop their analyses and simulations of flow instabilities. The study is the first attempt to present an analytical solution for this assumption.